\documentclass[12pt]{iopart}
\usepackage{graphicx}
\usepackage{array}
\usepackage{enumerate}
\begin{document}

\title[The inception of pulsed discharges above and below breakdown]%
{The inception of pulsed discharges in air: simulations in background
  fields above and below breakdown}

\author{Anbang Sun$^{1}$, Jannis Teunissen$^1$, Ute Ebert$^{1,2}$}

\address{$^1$Centrum Wiskunde \& Informatica (CWI), P.O.
  Box 94079, 1090 GB Amsterdam, The Netherlands,} \address{$^2$Dept.\
  Physics, Eindhoven Univ.\ Techn., The Netherlands,}
\ead{a.sun@cwi.nl}

\begin{abstract}
  We investigate discharge inception in air, in uniform background
  electric fields above and below the breakdown threshold.
  We perform 3D particle simulations that include a natural level of
  background ionization in the form of positive and O$_{2}^-$ ions.
  When the electric field rises above the breakdown and the detachment
  threshold, which are similar in air, electrons can detach from
  O$_{2}^-$ and start ionization avalanches.
  These avalanches together create one large discharge, in contrast to
  the `double-headed' streamers found in many fluid simulations.

  On the other hand, in background fields below breakdown, something
  must enhance the field sufficiently for a streamer to form.
  We use a strongly ionized seed of electrons and positive ions for
  this, with which we observe the growth of positive streamers.
  Negative streamers were not observed.
  Below breakdown, the inclusion of electron detachment does not
  change the results much, and we observe similar discharge
  development as in fluid simulations.
\end{abstract}

\maketitle

\section{Introduction}
\label{sec:introduction}
Developments in pulsed power technology have increased the interest in
pulsed discharges over the last two decades.
These discharges now have a wide range of applications, for example,
ozone generation \cite{samaranayake,ryo,vanVeldhuizen,heesch}, gas and
water cleaning \cite{vanVeldhuizen, grabowski,winands}, flow control
and plasma assisted ignition and combustion \cite{Starikovskiy2013}.
Pulsed discharges appear also in thunderstorms and in high voltage
technology for electricity networks.

Here, we focus on the initial development of such pulsed discharges in
air at standard temperature and pressure.
We consider two different cases for the background electric field.
It is either globally above the breakdown threshold, or only locally
due to some field enhancement.
In the first case, ionization processes can take place in the whole
volume.
In the second case, the discharge grows only in the region above
breakdown, forming a streamer discharge.
Streamers are fast growing plasma filaments that can penetrate into
non-ionized regions due the electric field enhancement at their tips.
In addition, streamers play an essential role in natural discharges,
since they pave the path for lightning and sprites.
They have been studied in different gases and in different electric
field configurations both experimentally
\cite{Nijdam2010,Nijdam2011a,veldhuizen2002,veldhuizen2003,yi,dubrovin,hegeler}
and numerically
\cite{Wormeester,Celestin2009,Tholin,Yair,luque2008,luque2008b,Aleksandrov,Naidis,Babaeva,dhali85}.

The main objective of the current paper is to show that in air one
needs to distinguish between background electric fields above and
below the breakdown threshold.
We will argue that an important reason for this distinction is the
presence of background ionization.
In atmospheric air, some background ionization is always present, due
to radioactivity, cosmic and solar radiation.
The free electrons that are generated quickly attach to molecules, to
form negative ions.
In dry air at a pressure of 1 bar, which we consider in this article,
the dominant negative ion is $\mathrm O_2^{-}$.
If the electric field rises above a threshold, called the detachment
field, electrons can again detach from these $\mathrm O_2^{-}$ ions
\cite{kossyi}.
Remarkably, the detachment field is very similar to the electrical
breakdown field in air.
So if the background field rises above the breakdown threshold,
electrons can detach, and every detached electron can start an
electron avalanche.

We want to emphasize that both detachment and photoionization are
characteristic for nitrogen/oxygen mixtures.
In pure gases or other mixtures they might be absent or much weaker,
see e.g.
\cite{Nijdam2010}.
Furthermore, in air at pressures below~100 mbar, $\mathrm O^{-}$ ions
become more important.
From these ions electrons can detach in fields much below
breakdown~\cite{gordillo2008,luquevazquez12,liu2012}.

We consider the inception of pulsed discharges in homogeneous
background fields, far from electrodes or other charge accumulations.
Such situations occur for example in thunderclouds
\cite{gurevich,liJiangbo2012}.
In our recent Geophysical Research Letter~\cite{sun}, we have used a
3D particle model to compare discharge formation in air at standard
temperature and pressure with and without natural background
ionization, in electric fields above the breakdown value.
We also briefly introduced an analytic estimate for the `ionization
screening time', after which the electric field in the interior of a
discharge is screened.
In the present paper, we further elaborate on the contents of the
letter \cite{sun}, and focus on the distinction between background
fields above and below breakdown.
Therefore, we present the evolution of discharges in fields both above
and below the breakdown threshold, using the same simulation model as
in~\cite{sun}.

The outline of the paper is as follows.
In Section~\ref{sec:model}, we introduce the simulation model, and
discuss electron detachment from background ionization.
In Section~\ref{sec:discharge_overvolted}, we present simulation
results for a background electric field above the breakdown threshold.
The formation of streamers in electric fields below the breakdown threshold is investigated in
Section~\ref{sec:streamer_undervolted}.  We there test what kind of initial ionization can lead to the
field enhancement required to start a streamer discharge.

\section{The set-up of the MC particle model}
\label{sec:model}
In recent years, we have developed a 3D particle code of the PIC-MCC
type \cite{birdsall} to study discharge inception.
The reason for using a 3D particle model is that the start of
discharges is often a stochastic process, that lacks cylindrical (or
other) symmetry.
In the model, electrons are tracked as particles.
Ions are assumed to be immobile, and are included as densities.
They only contribute to space charge effects.
Neutral gas molecules provide a background that electrons can randomly
collide with; they are included in the code as a random background of
given density.

The simulations of the present paper are performed in dry air (80\%
$\mathrm{N}_2$, 20\% $\mathrm{O}_2$) at 1 bar and 293 Kelvin.
For the electrons, we include elastic, excitation, ionization and
attachment collisions with the neutral gas molecules.
We use the cross sections from the SIGLO database \cite{pitchford} and
the null-collision method to select collisions \cite{nanbu}, with
isotropic scattering after every collision.
We ignore electron-electron and electron-ion collisions, because the
degree of ionization in a pulsed discharge in STP air is typically
below $10^{-4}$, which is also the case in the simulations we perform.

Simulating a discharge with a 3D particle code is computationally
expensive, especially as the discharge grows.
This limits the simulations we can perform to the first nanoseconds of
a discharge, during which the inception takes place.
On this time scale, heating, recombination and multi-step excitation
or ionization can be neglected.

\subsection{Adaptive particle management}
As the number of electrons in a typical discharge quickly rises to
$10^8$ or more, so-called super-particles have to be used.
Using super-particles with a fixed weight would induce significant
stochastic errors, and therefore we employ `adaptive particle
management' as described in~\cite{teunissen}.
The weight of simulated particles can then be adjusted by merging or
splitting them, and care is taken to not alter their properties in a
systematic way.
A particle $i$ can only be merged with its closest neighbor $j$ that
also needs to be merged, with `closest' defined as minimizing
\begin{equation}
  \label{eq:merge_distance}
  d^2 = (\vec{x}_i - \vec{x}_j)^2 + \lambda^2 \left|v_i - v_j\right|^2,
\end{equation}
where $\vec{x}$ denotes the Cartesian position vector, $v$ is the norm
of the velocity and $\lambda$ is a scaling factor that we set to one
picosecond.
A newly formed merged particle gets its velocity at random from one of
the original particles, while its position is set to the weighted
average position, see~\cite{teunissen} for a comparison of different
schemes to merge particles.
We adjust the weights so that every cell of the grid (see below)
contains at least 50 simulation particles.
So if no more than 50 electrons are present in a cell, then each
simulation particle represents a single electron.
But where the electron density is high, with much more than 50
electrons in a cell, most simulation particles represent many
electrons.

\subsection{Adaptive Mesh Refinement for the electric field}
In the particle code, the electric field is computed from the electric
potential.
The potential is computed by solving Poisson's equation with the
charge density as the source term, using the HW3CRT solver from the
FISHPACK library \cite{adams}.
When space charge effects become important in a discharge, a grid fine
enough to resolve the space charge structures has to be used.
For streamer discharges, that are surrounded by a thin space charge
layer, this means that a fine grid is required around the layer.
In our simulations, we use the following criterion for the grid
spacing
\begin{equation}
  \label{eq:refinement}
  \Delta x < 1/\alpha(E),
\end{equation}
where $\alpha(E)$ is the ionization coefficient, that describes the
average number of ionizations a single electron will generate per unit
length in a field of strength $E$.
For air at 1 bar and in an electric field of 15 MV/m, a typical field
for streamer tips, this gives $\Delta x \sim 5\;\mu\mathrm{m}$.
Because a typical simulation domain measures at least a few mm in each
direction, using such a fine grid everywhere is computationally
infeasible.
Therefore, we have implemented block-based adaptive mesh refinement,
in the same way as in~\cite{montijn06}, although now in 3D.
First, the electric potential is computed on a uniform, coarse grid.
Then the rectangular area that contains the points at which the
electric field is larger than some threshold is refined, by a factor
of two.
The electric potential in the refined rectangle is then computed by
imposing Dirichlet boundary conditions interpolated from the coarse
grid.
This procedure is repeated with the refinement criterion given by
equation~(\ref{eq:refinement}).

For the simulation of streamer discharges, the block-based grid
refinement strategy described above works relatively well, because
high electric fields are present only in a small region.
But for the simulation of discharges that spread out over the whole
domain, as we will see in section~\ref{sec:discharge_overvolted}, this
type of grid refinement does not reduce the computational cost much.

\subsection{Photoionization}
Photoionization provides a non-local ionization mechanism in air.
This is especially important for the propagation of positive
streamers, that need a source of free electrons ahead of them to
propagate.
We use the same approach as in \cite{Chanrion08} and \cite{li2011},
where a discrete, stochastic version of Zhelezniak's photoionization
model~\cite{Zhelezniak} is implemented.
In this model, the average density of ionizing photons $S_\mathrm{ph}$
produced at $\vec{r}$ is given by
\begin{equation}
  \label{eq:photoionization}
  S_\mathrm{ph}(\vec{r}) = S_\mathrm{ion}(\vec{r}) \; \eta(E),
\end{equation}
where $S_\mathrm{ion}$ represents the number of ionizations and
$\eta(E)$ is an efficiency, estimated from experimental measurements,
that depends on the local electric field $E$ and the gas mixture.
When an ionizing photon is generated, its place of absorption is
determined using random numbers, and at that position an electron-ion
pair is created.
The average absorption distance is about $0.5$ mm in air at 1 bar.
For details about the implementation of the photoionization model we
refer to \cite{Chanrion08}.

\subsection{Electron detachment from background ionization}
\label{sec:detachment}

\begin{figure}
  \centering
  \includegraphics[width=0.6\textwidth]{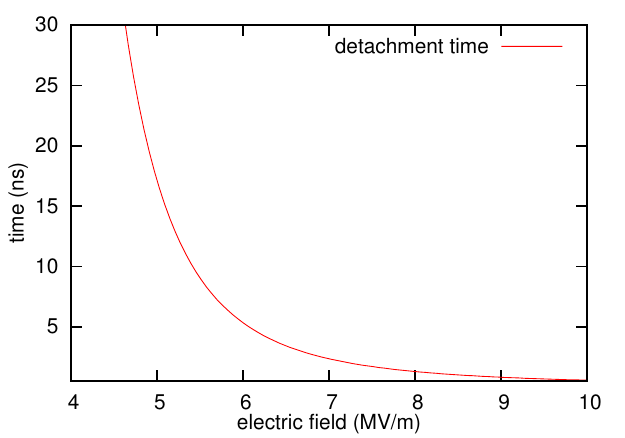}
  \caption{The detachment time $\tau_D$ as a function of the electric
    field strength in STP air.
    In higher fields, negative ions have a higher energy and drift
    faster, so they are more likely to lose an electron in a collision
    with a neutral molecule.}
  \label{fig:detachment_time}
\end{figure}

In atmospheric air, there is always some background ionization
present, due to radioactivity and cosmic or solar radiation.
Previous discharges can also play a role, both in nature
\cite{luquevazquez2011} and in the lab \cite{Nijdam2011a}.
At standard temperature and pressure, the free electrons that are
created by these sources attach to oxygen molecules mostly by
three-body attachment~\cite{kossyi}:
\begin{eqnarray}
  \mathrm{e}+\mathrm{O}_2+\mathrm{O}_2 &\rightarrow \mathrm{O}_2^{-} + \mathrm{O}_2,
  \label{eq:att1}\\
  \mathrm{e}+\mathrm{N}_2+\mathrm{O}_2 &\rightarrow \mathrm{O}_2^{-} + \mathrm{N}_2.
  \label{eq:att2}
\end{eqnarray}
These negative ions have a longer life time than the electrons.
Inside buildings, background ionization levels of $10^3$ - $10^4 \;
\mathrm{cm}^{-3}$ are typical, primarily due to the decay of radon,
see~\cite{pancheshnyi05} for a review.
When $\mathrm O_2^{-}$ molecules collide with a neutral gas molecule,
they can lose an electron.
This can be regarded as the inverse of the reactions in Equations
(\ref{eq:att1}) and (\ref{eq:att2}):
\begin{eqnarray}
  \mathrm{O}_2^{-} + \mathrm{O}_2 &\rightarrow \mathrm{e}+\mathrm{O}_2+\mathrm{O}_2,
  \label{eq:det1}\\
  \mathrm{O}_2^{-} + \mathrm{N}_2 &\rightarrow \mathrm{e}+\mathrm{N}_2+\mathrm{O}_2.
  \label{eq:det2}
\end{eqnarray}
The rate constants for these detachment reactions can be related to
the reduced electric field $E/N$ and the number density of the
neutrals.
We use the rates given in \cite{kossyi}.
For a given number density of the neutrals, we call the total rate at
which electrons detach from $\mathrm{O}_2^{-}$ ions the
\emph{detachment rate}.
Additionally, we call the inverse of the detachment rate the
detachment time $\tau_D$.
In figure~\ref{fig:detachment_time}, the dependence of $\tau_D$ on the
electric field strength is shown.

If during a simulation an electron has an attachment collision, then
the electron is removed and an $\mathrm{O}_2^{-}$ ion is created at
the electron's position.
We currently consider only this type of negative ion, although
$\mathrm{O}^{-}$ can also form due to dissociative attachment, mostly
at lower pressures or higher electron energies, and many more ions can
be generated by chemical reactions \cite{luquevazquez12, popov}.

\section{Discharges in background fields above breakdown}
\label{sec:discharge_overvolted}

\subsection{Previous work}

Up to now, pulsed discharges in air have mainly been simulated with
plasma fluid models
\cite{luque2008,pasko1998,liu04,qin2012,pasko2007,bourdron2007,bessieres},
where the charged particles are approximated by densities.
The most common fluid model assumes that the electrons drift, diffuse
and react (ionize), with the coefficients for these processes
determined by the local electric field strength.
Typically cylindrical symmetry is assumed, and therefore these fluid
models need just two spatial coordinates, making them computationally
much less expensive than our 3D particle code.
Authors typically place some localized initial ionization in the
domain to start a discharge
\cite{luque2008,pasko2007,bourdron2007,bessieres}.
In background fields above the breakdown threshold, this ionization
seed then develops into a `double-headed' streamer, where both the
positive and the negative end grow simultaneously.
The effect of including natural background ionization (and detachment)
has not been studied with these models.

However, including background ionization and detachment is very
important for discharges in air above breakdown, as we have recently
demonstrated in \cite{sun}.
There we presented 3D particle simulations in a background field of
7~MV/m (where the breakdown field is 3~MV/m) with three different
initial conditions: either only one electron, or one electron together
with a background density of 10$^3$ O$_2^-$ ions per cm$^3$, or this
background density alone without an initial electron.
Our results showed that in the first case indeed a double-headed
streamer emerged, while in both cases with a realistic background ion
density, the discharge developed in a much more homogeneous manner:
The natural background ionization together with the detachment
reaction generates free electrons everywhere in the region above
breakdown.
As this is a stochastic process, the resulting discharge does not have
cylindrical symmetry, which is why we use the 3D particle model
introduced in section~\ref{sec:model}.
Simulations with a background density of 10$^4$ O$_2^-$ ions per
cm$^3$ in background fields of 6~MV/m are presented in
section~\ref{sec:3dresults_OV}.

\subsection{Boundary conditions for the simulations above breakdown}
\label{sec:effect_BG_overVolted}

\begin{figure}
  \centering
  \includegraphics[width=0.2\textwidth]{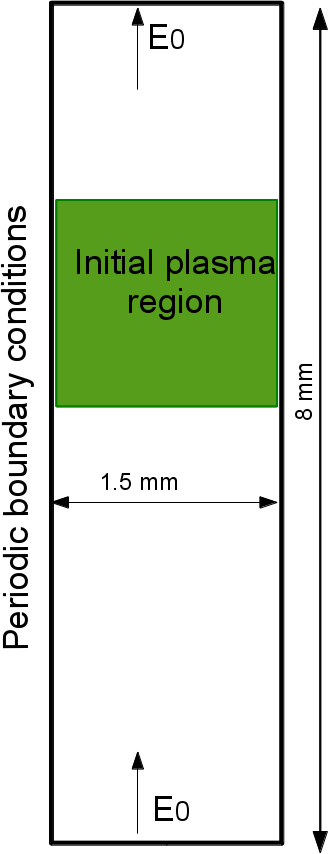}
  \caption{Schematic of the computational domain in the 3D overvolted
    simulations.
    Periodic boundary conditions are used in the two lateral
    directions.
    At the top and bottom of the domain, the electric field is fixed
    to a value $E_0$ of 6~MV/m.
    Initially, background ionization is present in the green region.}
  \label{fig:computational_domain_overvolted}
\end{figure}

In this section, we present new simulation results for a discharge
developing from a natural level of background ionization of O$^-_2$
ions in a field of 6~MV/m, well above breakdown.
We want to simulate the development of a discharge that is not in
contact with physical boundaries, like electrodes.
This is achieved by using periodic boundary conditions in the $x$ and
$y$ direction, while limiting the region where background ionization
is present in the $z$ direction.
In other words, we simulate the development of a thick discharge layer
growing from background ionization.
The elongated computational domain is shown in
Figure~\ref{fig:computational_domain_overvolted}, where the region
with background ionization is shaded green.
At the top and bottom of the domain we apply Neumann boundary
conditions for the electric potential, thereby creating a uniform
background field $E_0$ of 6~MV/m.

We remark that in the GRL~\cite{sun} we were less careful with the
boundary conditions and used something similar to
Fig.~\ref{fig:computational_domain_undervolted}.
As the complete pre-ionized region becomes electrically screened, the
boundary of the pre-ionized region induced distortions of the electric
field.

We do not use grid refinement to calculate the generated electric
field in this simulation as grid refinement would be required nearly
everywhere in the pre-ionized region.
The static grid contains $100 \times 100 \times 535$ cells, with a
cell length of $15 \; \rm {\mu m}$.
The domain length is chosen in such a manner that the discharge does
not reach its boundaries within the time
simulated.

\begin{figure}
  \begin{center}
    \noindent\includegraphics[width=0.9\textwidth]{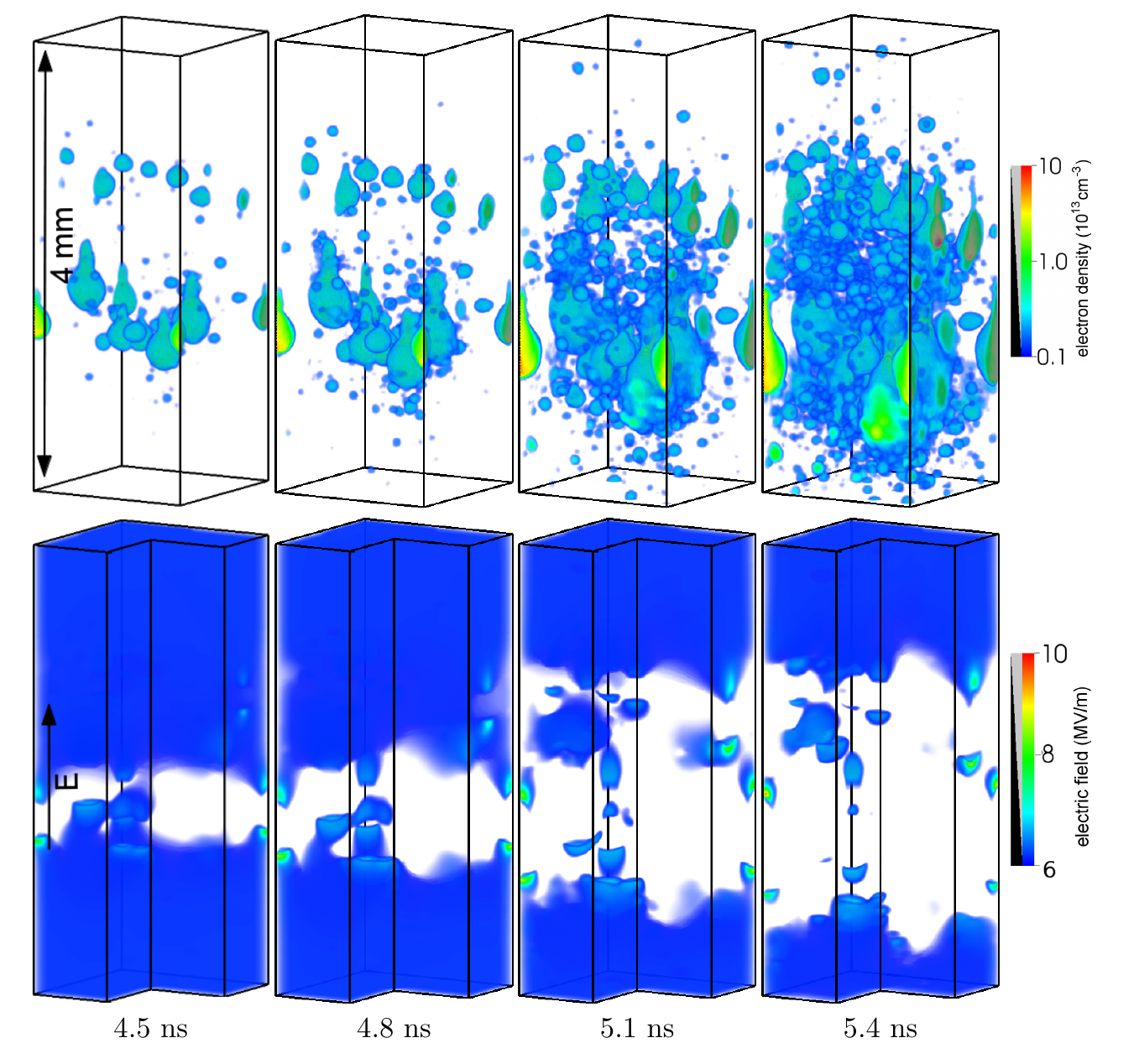}
    \caption{The time evolution of the electron density (top row) and
      of the electric field (bottom row).
      Background ionization is initially present in the green region
      of Figure~\ref{fig:computational_domain_overvolted}, in the form
      of $\mathrm{O}_2^-$ and positive ions, both with a density of
      $10^{4}\;\mathrm{cm}^{-3}$.
      The gas is dry air at 1 bar and 293 K in an upward directed
      homogeneous electric field of 6 MV/m, which is about two times
      the breakdown field.
      The domain between 2 mm and 6 mm in the vertical direction of
      Figure~\ref{fig:computational_domain_overvolted} is shown.
      The figures were generated using volume rendering, and the
      opacity is shown next to the colorbar; black indicates
      transparency.
      For the figures in the second row, a quarter of the domain is
      removed to show the inner structures of the electric field.
    }
    \label{fig:from_detachment_overvolted}
  \end{center}
\end{figure}

\subsection{Simulated discharge evolution}
\label{sec:3dresults_OV}

Figure~\ref{fig:from_detachment_overvolted} shows the evolution of the
electron density and the electric field in four time steps between 4.5
ns and 5.4 ns.
The evolution of the discharge can be characterized as follows.
First, free electrons appear due to detachment.
As can be seen in figure~\ref{fig:detachment_time}, the characteristic
detachment time in a field of 6 MV/m is about 5 ns.
Then these free electrons start electron avalanches, that quickly grow
due to impact ionization.
The growing avalanches also produce photoionization, thereby starting
additional avalanches.
Eventually, many avalanches emerge in the simulation domain.

\begin{figure}
  \begin{center}
    \noindent\includegraphics[width=0.45\textwidth]{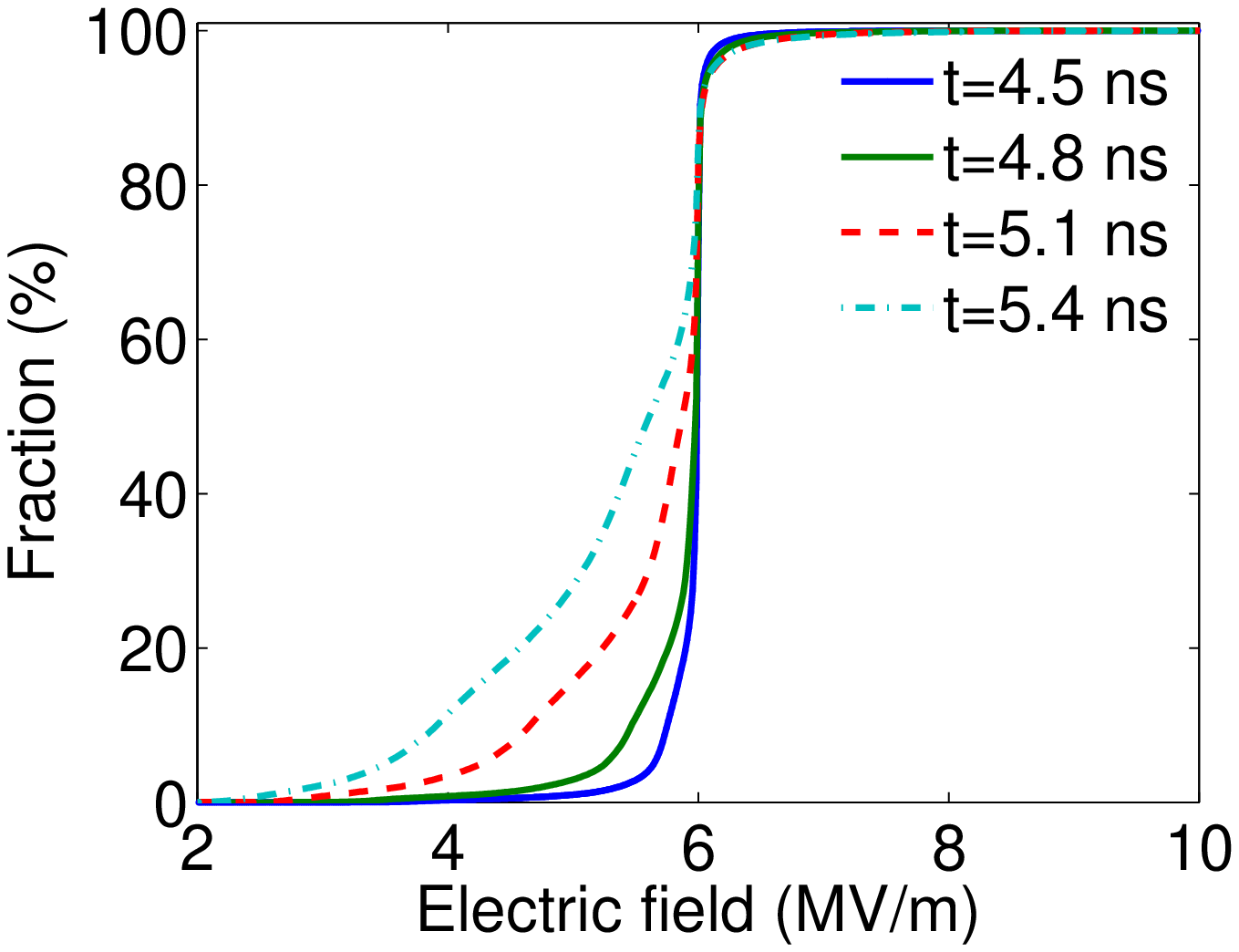}
    \caption{The volume fraction with a field smaller than $E$ as a
      function of $E$.
      We evaluate the green domain shown in
      figure~\ref{fig:from_detachment_overvolted}.}
    \label{fig:fraction_efield}
  \end{center}
\end{figure}

After about 5 ns, space charge effects start to become important,
causing the electric field to increase locally up to $\sim$ 9 MV/m
while decreasing elsewhere.
These space charge effects increase in magnitude until the simulation
stops at 5.4 ns.
The distribution of the electric field values is shown in
figure~\ref{fig:fraction_efield}.
After 4.5~ns almost the complete system is still at the background
field of 6~MV/m, while about 8$\%$ of the volume has a field lower
than the breakdown value of 3~MV/m after 5.4~ns.

\begin{figure}
  \begin{center}
    \noindent\includegraphics[width=0.45\textwidth]{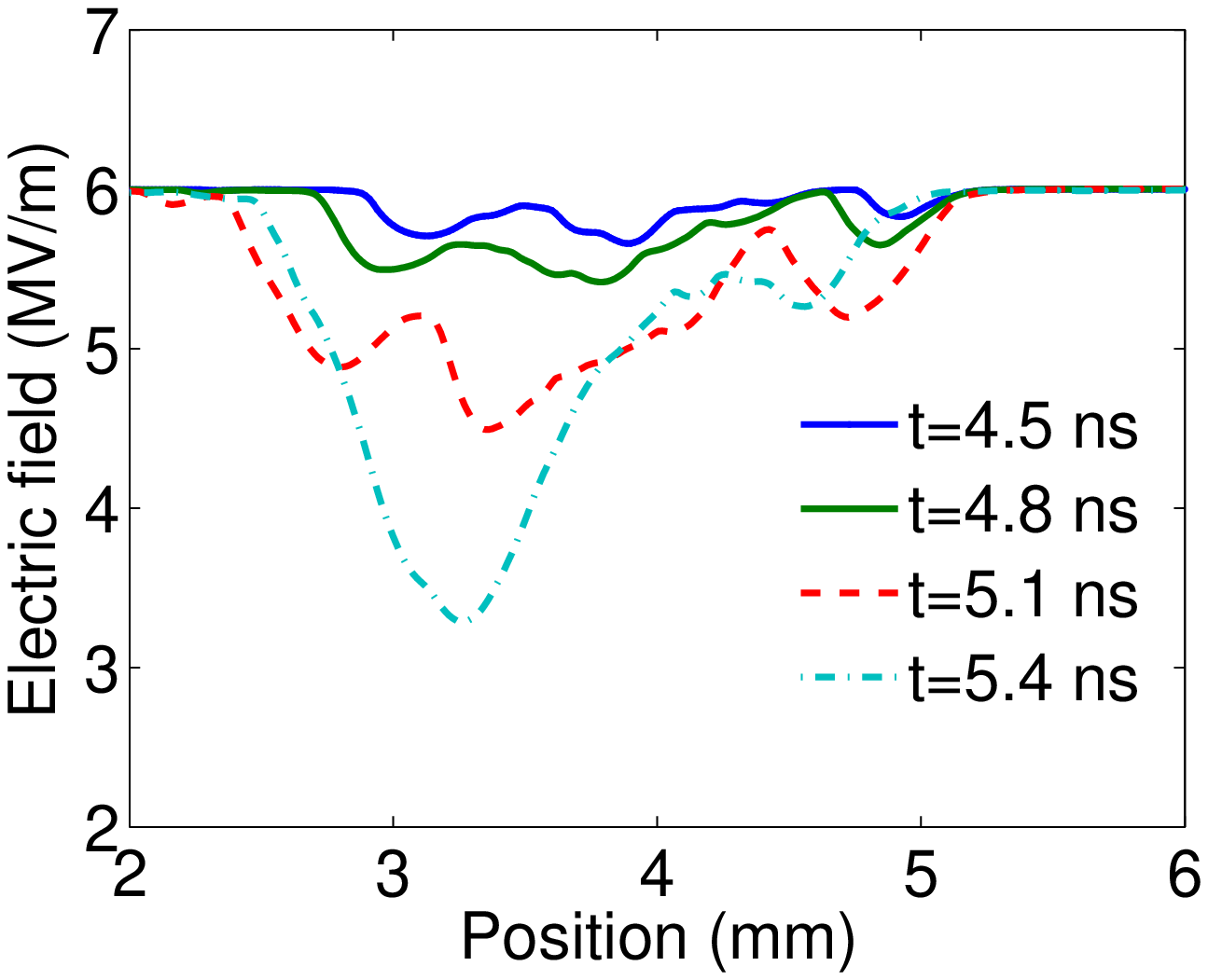}
    \caption{The screened electric field at different times.
      The electric field was averaged over planes perpendicular to the
      background field in Figure~\ref{fig:from_detachment_overvolted}
      and plotted as a function of the coordinate parallel to the
      field.}
    \label{fig:efield_distribution_NegO2_6MV}
  \end{center}
\end{figure}

Figure~\ref{fig:efield_distribution_NegO2_6MV} shows the distribution
of electric fields in the simulation in another manner; it shows the
electric field averaged over the horizontal planes intersecting
Figure~\ref{fig:from_detachment_overvolted} and plotted as a function
of the longitudinal coordinate.
The screening of the electric field occurs in a `noisy' way, and the
electric field varies significantly inside the discharge.
This is not so surprising, as initially only about 45 negative ions
($\mathrm{O}_2^-$) are present.
With these ions randomly placed in a volume of 4.5 mm$^3$, we do not
expect a discharge homogeneously filled with ionization.

The simulation stops when there are too many simulation particles for
the computer's memory, which happened here at about $3\cdot 10^7$
particles.

\subsection{Conclusions and further work}

Two conclusions can be drawn from these results:
\\
First, the simulations show a rather global breakdown in the
pre-ionized region, instead of a double-headed streamer.
This is due to the inclusion of a natural background of oxygen ions
and the electron detachment reaction from these ions in a field
approximately above the breakdown value.
\\
Second, the breakdown is not completely uniform either, but shows a
competition of local streamer formation and global breakdown that
creates a certain patchiness of ionization and makes the electric
field screening noisy.

A detailed analysis of these phenomena and their relation to the
concept of an ionization screening time~\cite{sun} will be discussed
in a forthcoming paper~\cite{newScreeningPaper}.

\section{Discharges in background fields below breakdown}
\label{sec:streamer_undervolted}

\subsection{Previous work}

In this section, we investigate streamer formation in background
electric fields below the breakdown threshold.
Will the 3D particle model together with background ionization and
electron detachment also lead to major deviations from previous
results derived with a 2D fluid model?

In such a simulation, of course, the field has to exceed the breakdown
threshold in some region, otherwise a discharge cannot start.
An electric field that is only locally above breakdown can be
generated in several ways.
In experiments, such a field can be created by sharp electrodes, to
which a voltage is applied.
Another possibility is that a conducting or polarizable object floats
in a field below breakdown; examples include dust, water droplets or
ice crystals.
At the endpoints of the object, the field will increase, especially if
the object has an elongated shape along the direction of the
background field.
Yet another way to start a discharge is to have a strongly ionized
region that therefore acts as a conductor.
This typically requires electrons, as they have a much higher mobility
than ions.

Such ionized seeds have been commonly used in fluid simulations of
streamers for the past 30 years, often without explicit mentioning.
Liu \textit{et al} \cite{liu12a} used an ionized column as a
substitute of a droplet or ice particle to start streamer discharges.
Assuming that the ionized column was perfectly conducting and had a
typical streamer radius, they presented an estimate on its minimal
length to provide sufficient field enhancement \cite{liu12a,
  bazelyan}.
A positive streamer was able to form if the electric field at the tip
can be enhanced to $\sim 3 - 5 E_k$, where $E_k$ is the breakdown
field of air at standard temperature and pressure.
To generate significant field enhancement, the density of an initial
seed should be comparable to the density inside a streamer channel,
which is $10^{12} - 10^{14}\;\mathrm{cm^{-3}}$ in atmospheric air.
In \cite{kosar}, the possible sources of such strong preionization at
70 km altitude are discussed.
It is still an open question where these seeds would come from in
atmospheric air, for example to start lightning discharges
\cite{liu12a}.

Below we first present two examples of seeds that enhance the field
sufficiently for positive streamers to form in our particle model.
Then we investigate the effect of electron detachment and natural
background ionization on the formation process.
We again use dry air at 1 bar and 293 Kelvin with a density of 10$^4$
ions per cm$^3$.

\subsection{Boundary conditions for the simulations below breakdown}
\label{sec:undervolted}

\begin{figure}
  \centering
  \includegraphics[width=0.3\textwidth]{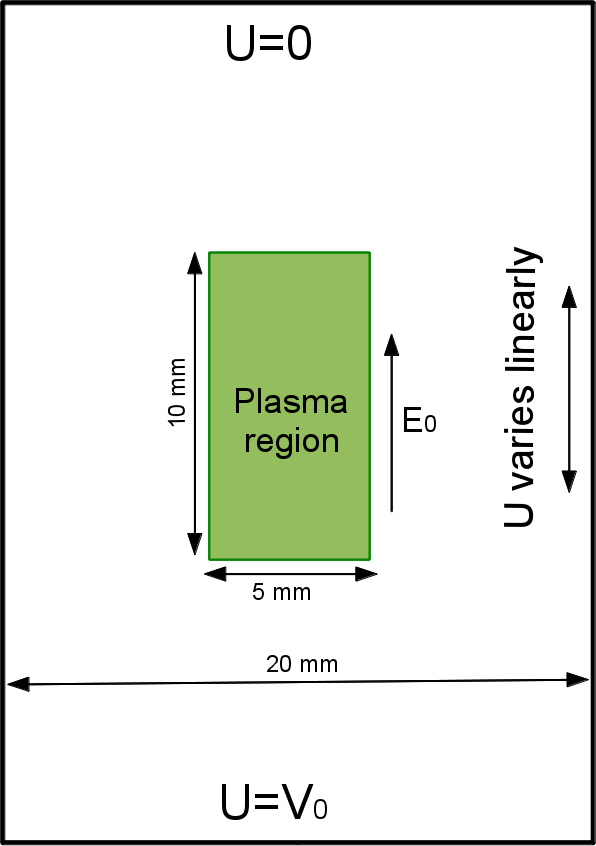}
  \caption{Schematic of the computational domain in the undervolted
    simulations.
    The simulated plasma region is embedded in a 4 times larger
    system.
    The potential of the large volume is calculated first, then the
    result is interpolated to get the boundary potential of the small
    domain.
    Dirichlet boundary conditions are used at all the boundaries.}
  \label{fig:computational_domain_undervolted}
\end{figure}

The computational domain that we use for the simulations in background
fields below breakdown is shown in
Figure~\ref{fig:computational_domain_undervolted}.
Because we want to study the development of a single, isolated
streamer discharge, we cannot use periodic boundary conditions as we
did in section~\ref{sec:discharge_overvolted}.
Instead, an additional grid is introduced, to be able to set boundary
conditions for the electric potential farther away.
The complete computational domain thus has two parts: an interior grid
of $5\times 5 \times 10$ mm$^3$, in which we use the particle model,
and a four times larger grid around it that is used to set the
boundary conditions for the electric potential on the interior grid.
Dirichlet boundary conditions are imposed on the sides of the larger
grid to get a homogeneous background field $E_0 < E_k$ in the vertical
direction.
Inside the interior grid we use adaptive mesh refinement, so that the
strong electric fields around streamer heads can be resolved.

\subsection{Streamer inception from conductive seeds}

We present results for two initial seeds here, that are both placed at
the center of the domain.

\subsubsection{First seed}
\label{sec:UV_first_seed}
The first seed we use is a long, neutral ionized column.
The peak ion and electron density is
$1.3\times10^{13}\;\mathrm{cm}^{-3}$.
In the two lateral directions, the distribution of electrons and ions
is Gaussian, with a width of 0.2 mm.
The distribution of plasma in the vertical direction is uniform over a
length of 4 mm; at the endpoints there is again a Gaussian
distribution.
An external electric field of $\sim 0.5 E_{k}$ is applied in the
vertical direction.
This seed is similar to the initial condition that was used in a 2D
fluid model \cite{liu12a}, but then scaled to ground pressure.

Figure~\ref{fig:liu_inition_undervolted} shows how this seed develops
further in the simulations.
The ionized column rapidly gets polarized, because the electrons drift
against the electric field.
A negative and positive charge layer emerge at the top and bottom of
the column, respectively.
After $\sim 10$ ns, a positive streamer forms at the upper end of the
column, as shown in the first row of
Figure~\ref{fig:liu_inition_undervolted}.
At the lower end, electrons spread out or attach to neutral
molecules.
On the time scales that can be simulated with our particle model, we
have not observed negative streamers emerging.
An important difference between positive and negative streamers is
that positive streamers grow from electrons drifting inwards towards
their head, while negative streamers grow from the electrons drifting
outwards.
Thus, the space charge layer of a positive streamer head is formed by
rather immobile ions, while the space charge layer of a negative
streamer head is formed by mobile electrons.
Negative streamers are therefore typically wider and more diffusive,
with less field enhancement, and they do not form as easily
\cite{luque2008}.

\subsubsection{Second seed}
\label{sec:UV_second_seed}
The second seed has an isotropic Gaussian distribution for electrons
and ions in three spatial dimensions.
Such seeds have frequently been used to study the formation of
positive streamers in point-to-plane gaps
\cite{Kulikovsky,serdyuk,Bonaventura}.
The Gaussian distribution we use here has a peak density of $2.3
\times 10^{13}\;\mathrm {cm}^{-3}$ and a width of 0.3 mm.
A homogeneous electric field of $\sim 0.7 E_k$ is applied in the
upward direction.
to reduce the simulation time.

\begin{figure}
  \begin{center}
    \noindent\includegraphics[width=0.90\textwidth]{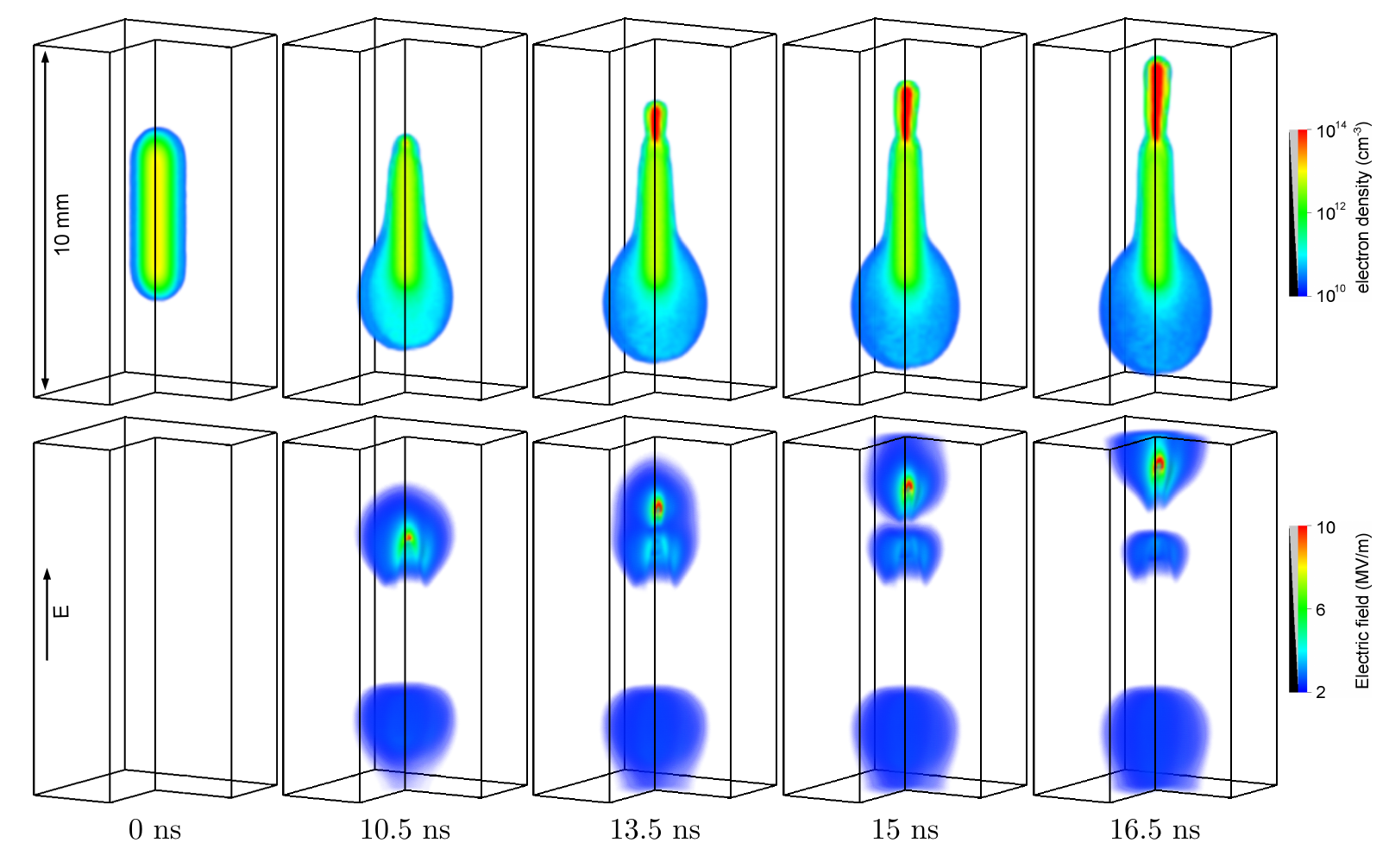}
    \caption{ The electron density (top row) and the electric field
      (bottom row) in a 3D particle simulation with photoionization
      and natural background ionization ($10^4\;\mathrm{cm}^{-3}$).
      Times are indicated below each column.
      The simulation starts from an ionized column, with a length of 4
      mm, a characteristic width of 0.2 mm and a peak plasma density
      of $1.3 \times 10^{13}\;\mathrm{cm}^{-3}$.
      The gas and plots were set up in the same way as that for
      Figure~\ref{fig:from_detachment_overvolted}, but in an upward
      homogeneous electric field of 1.7 MV/m ( about 0.5 times $E_k$).
      The simulation domain has a length of 10 mm in the vertical
      direction and of 5 mm in the two lateral directions.
    }
    \label{fig:liu_inition_undervolted}
  \end{center}
\end{figure}

\begin{figure}
  \begin{center}
    \noindent\includegraphics[width=0.90\textwidth]{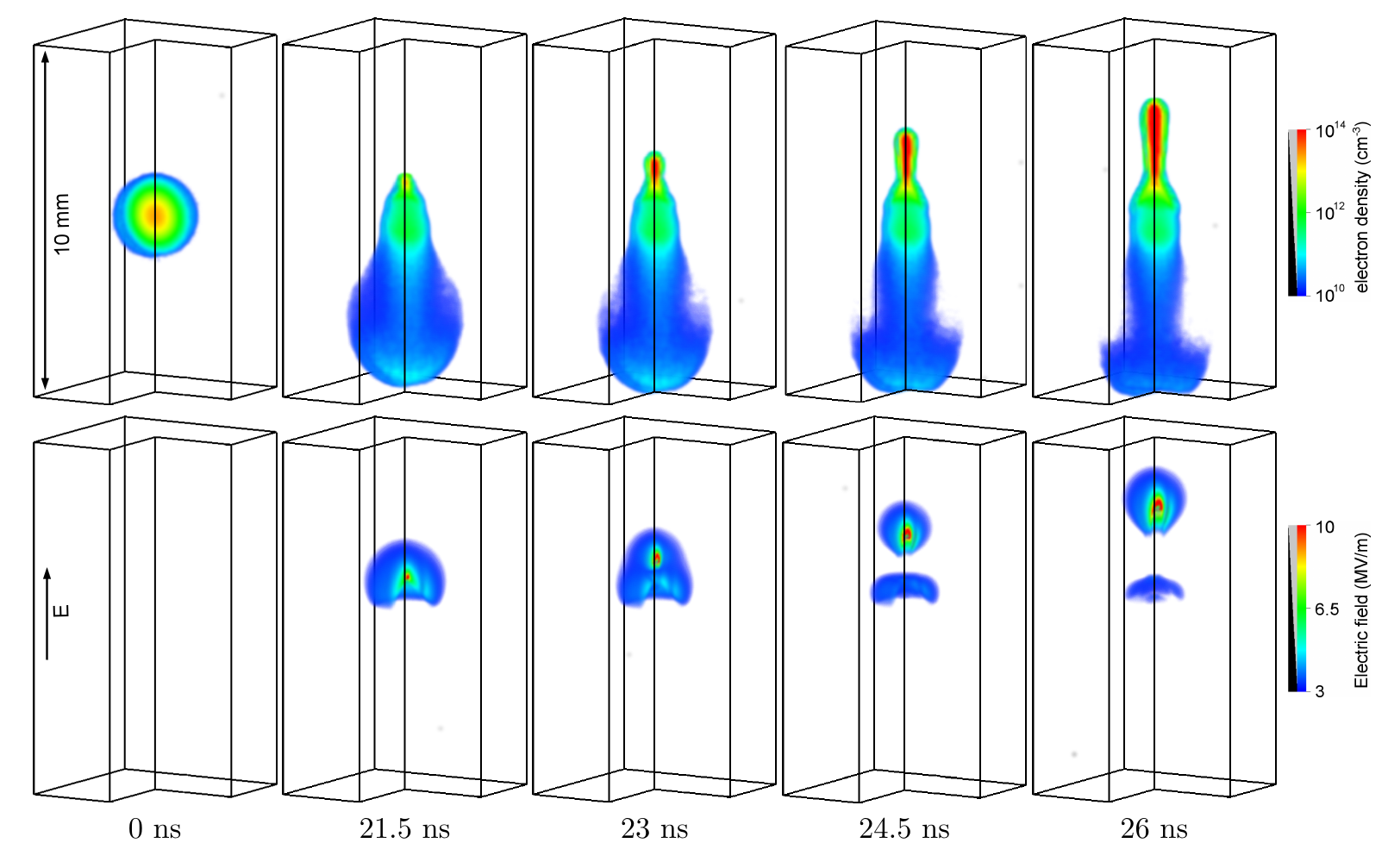}
    \caption{ The distributions of electron density (top row) and
      electric field (bottom row) in a 3D particle model with
      photoionization and natural background ionization
      ($10^4\;\mathrm{cm}^{-3}$).
      Times are indicated below each column.
      The simulation starts from a neutral Gaussian plasma seed, with
      a characteristic radius of 0.3 mm and a peak plasma density of
      $2.3 \times 10^{13}\;\mathrm{cm}^{-3}$.
      Gas, simulated domain and plots were set up in the same way as
      that in Figure~\ref{fig:liu_inition_undervolted}, but in an
      upward homogeneous electric field of 2.5 MV/m ( about 0.7 times
      $E_k$).
    }
    \label{fig:gaussian_inition_undervolted}
  \end{center}
\end{figure}

Figure \ref{fig:gaussian_inition_undervolted} shows the evolution of
the electron density and the electric field.
The development is similar to
figure~\ref{fig:liu_inition_undervolted}: the seed is rapidly
polarized and two charge layers form.
When the maximum electric field reaches $\sim 3 E_k$ at the upper tip,
a positive streamer emerges and propagates upward.
As before, we do not observe negative streamers on this time scale.

\subsubsection{Discussion}
Both seeds have an electron density comparable to a streamer channel,
so it is no surprise that a streamer can develop.
The elongated seed causes stronger field enhancement, and therefore
positive streamers form more easily than with the isotropic Gaussian
seed.
In general, the following can be said about such ionized seeds.
First, the seed has to be sufficiently conductive, so that the
electric field in its interior gets reduced, and the electric field at
its boundary gets enhanced.
The required electron density for this is approximately that of a
streamer.
Second, the longer the seed size is in the direction of the applied
electric field, the stronger the field enhancement is at the
endpoints.
For a conducting sphere, the enhancement factor over the background
field is three, but for more elongated shapes it can be much higher.
We did not systematically explore which seeds can cause streamer
inception under what conditions, because such simulations are very
time consuming with our 3D particle model.

\subsection{The role of detachment and natural background ionization
  for streamer inception}

\begin{figure}
  \begin{center}

    \noindent\includegraphics[width=0.5\textwidth]{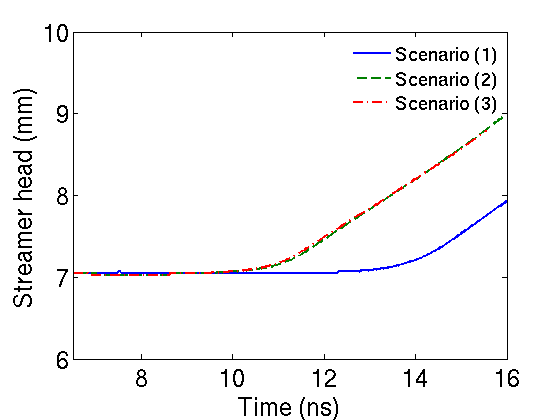}

    \caption{The vertical position of the streamer head as a function
      of time, see figure~\ref{fig:liu_inition_undervolted}.
      We define the position of the head as the location where
      electric field in vertical direction is maximal.
      The seed and the background electric field are the same as in
      Section \ref{sec:UV_first_seed}.
      Three scenarios are tested: (1) Without electron detachment, so
      only including photoionization; (2) With an initial
      $\mathrm{O}_2^{-}$ density of $10^4\;\mathrm{cm}^{-3}$; (3) With
      an initial $\mathrm{O}_2^{-}$ density of
      $10^7\;\mathrm{cm}^{-3}$.}
    \label{fig:streamer_position}
  \end{center}
\end{figure}

Above we have seen that an conductive seed can provide enough field
enhancement for positive streamers to start.
However, for the start of a discharge not only a high field is
required, but also some free electrons.
With the ionized seeds presented above, there are many free electrons
present in the simulation.
But the field enhancement at the positive side of these seeds happens
because electrons have moved away from there, so with just the seed it
is still hard to start a discharge.
Here, we investigate how electron detachment can provide the free
electrons to start a discharge.
We use the ionized column that was introduced in Section
\ref{sec:UV_first_seed} in a field of $0.5\; E_k$.
In Figure \ref{fig:streamer_position}, the position of the streamer
head is shown versus time, for three different scenarios:
\begin{enumerate}[(1)]
  \item Without electron detachment, including only photoionization
  \item With an initial $\mathrm{O}_2^{-}$ density of
  $10^4\;\mathrm{cm}^{-3}$
  \item With an initial $\mathrm{O}_2^{-}$ density of
  $10^7\;\mathrm{cm}^{-3}$
\end{enumerate}

We observe that positive streamers form a bit earlier if we include
the detachment process, but that the density of $\mathrm{O}_2^{-}$
molecules has little influence.
The reason is that the initial seed has an electron density orders of
magnitude higher than the $\mathrm{O}_2^{-}$ density.
Therefore, it is mostly electron attachment at the beginning of the
simulation, when the field is still low, that determines the actual
$\mathrm{O}_2^{-}$ density.
When the field has become large enough for detachment, the discharge
can start faster due to the extra detaching electrons.

On the other hand, we observe that electron detachment has no effect
on the propagation of the positive streamer.
The reason is that photoionization produces most free electrons ahead
of the front after the discharge has started~\cite{Wormeester}.

\subsection{Comparison with 2D fluid models}

In fields above the breakdown threshold, our 3D particle simulations
were very different from previous simulations with 2D plasma fluid
models.
An important factor for this difference was the inclusion of O$_2^{-}$
ions due to background ionization, from which electrons could detach.
On the other hand, in fields below breakdown, our results were in
agreement with typical 2D fluid simulations.
The reason is that electron impact ionization and electron detachment
occur almost only where the field is above breakdown.
When the background field is above breakdown, these processes can
happen anywhere, and a global discharge forms.
But when the background field is below breakdown, these processes
happen where the field is locally enhanced: at the streamer head.
Around the positive streamer head photoionization will typically be
the dominant source of free electrons, so background ionization and
electron detachment are not so important.

\section{Conclusion}

We have studied pulsed discharge formation in electric fields above
and below the breakdown threshold with a 3D particle model for air at
standard temperature and pressure.
Photoionization, a natural level of O$_2^{-}$ ions due to background
ionization and electron detachment were included.

In background electric fields above breakdown, we see discharges
distributed over the whole domain instead of the `double-headed'
streamers often appearing in other publications.
The major cause for this difference is the inclusion of background
ionization and detachment.
Free electrons appear at many different places due to detachment from
O$_2^-$ ions, and start electron avalanches.
These avalanches interact and overlap, and can eventually screen the
electric field in the interior of the discharge.
This process is analyzed in a separate paper~\cite{newScreeningPaper}.

In background electric fields below the breakdown value, positive
streamers can form if the field is locally sufficiently enhanced.
We have shown two examples in which an conductive seed of electrons
and ions causes sufficient field enhancement for a positive streamer
to grow.
Negative streamers do not appear in our particle simulations.
In fields below breakdown, our 3D particle model gives similar results
as the 2D plasma fluid models used by other authors.
The reason is that electron detachment and impact ionization only
occur around the streamer head, where the field is enhanced.
Since photoionization is the dominant source of free electrons around
the streamer head, electron attachmennt and detachment are only
important during the inception phase of the discharge.

\ack{ABS acknowledges the support by an NWO Valorization project at
  CWI and by STW projects 10118 and 12119.
  JT was supported by STW project 10755.}

\end{document}